\documentclass[aps,prd,preprint,preprintnumbers,nofootinbib,superscriptaddress]{revtex4}
\usepackage{ae}
\usepackage{bm} 
\usepackage{color}
\usepackage{amsmath}
\usepackage{amssymb}
\usepackage{amsfonts}
\usepackage{graphicx}
\usepackage{slashed}
\usepackage{setspace}
\usepackage{stackrel}
\usepackage{ulem}
\usepackage{color}

\newcommand{\Eqref}[1]{Eq.~\eqref{#1}}

\allowdisplaybreaks

\begin{document}

\setlength{\unitlength}{1mm}
\title{Vacuum birefringence in the head-on collision of XFEL and optical high-intensity laser pulses}
\author{Felix Karbstein}\email{felix.karbstein@uni-jena.de}
\affiliation{Helmholtz-Institut Jena, Fr\"obelstieg 3, 07743 Jena, Germany}
\affiliation{Theoretisch-Physikalisches Institut, Abbe Center of Photonics, \\ Friedrich-Schiller-Universit\"at Jena, Max-Wien-Platz 1, 07743 Jena, Germany}
\date{\today}

\begin{abstract}
 The focus of this article is on providing compact analytical expressions for the differential number of polarization flipped signal photons constituting the signal of vacuum birefringence in the head-on collision of x-ray free electron (XFEL) and optical high-intensity laser pulses.
 Our results allow for unprecedented insights into the scaling of the effect with the waists and pulse durations of both laser beams, the Rayleigh range of the high-intensity beam, as well as transverse and longitudinal offsets.
 They account for the decay of the differential number of signal photons in the far-field as a function of the azimuthal angle measured relative to the beam axis of the probe beam in forward direction, typically neglected by conventional approximations. Moreover, they even allow us to extract an analytical expression for the angular divergence of the perpendicularly polarized signal photons.
 We expect our formulas to be very useful for the planning and optimization of experimental scenarios aiming at the detection of vacuum birefringence in XFEL/high-intensity laser setups, such as the one put forward at the Helmholtz International Beamline for Extreme Fields (HIBEF) at the European XFEL.  
\end{abstract}

\maketitle

\section{Introduction}\label{sec:intro}

In contrast to classical notions of the vacuum, the vacuum of relativistic quantum field theories such as quantum electrodynamics (QED), describing the microscopic interactions of light and matter, is far from trivial: the reason being the omnipresence of fluctuations of the theory's quantum fields. 
While QED does not provide for a direct tree-level interaction among photons, quantum vacuum fluctuations generically induce effective nonlinear couplings among electromagnetic fields \cite{Euler:1935zz,Heisenberg:1935qt,Weisskopf,Schwinger:1951nm}, the leading one arising from an effective four-photon coupling mediated by an electron-positron loop \cite{Karplus:1950zza,Costantini:1971cj}.
For the macroscopic electromagnetic fields available in the laboratory, these effective couplings are very small and suppressed parametrically with the electron mass $m$, constituting the typical scale the strengths of the applied fields are compared to; note that $mc^2\simeq 511\,{\rm keV}$, corresponding to very large magnetic and electric field strengths of $\frac{m^2c^2}{e\hbar}\simeq4.4\cdot10^9\,{\rm T}$ and $\frac{m^2c^3}{e\hbar}\simeq1.3\cdot10^{16}\,\frac{\rm V}{\rm cm}$, respectively.
This explains why signatures of QED vacuum nonlinearity are very elusive in experiment, and have not been verified experimentally in macroscopic electromagnetic fields so far.

One of the most fascinating properties of the QED vacuum in strong electromagnetic fields is vacuum birefringence \cite{Toll:1952,Baier,BialynickaBirula:1970vy,Adler:1971wn}.
If an originally linearly polarized probe photon beam is sent through a strong-field region, some of the photons constituting the beam can be scattered into a perpendicularly polarized mode, whereas the majority
of the probe photons traverses the strong-field unaltered. This effectively supplements the probe beam with a tiny ellipticity, thereby attributing a birefringence phenomenon to the quantum vacuum in electromagnetic fields.

Vacuum birefringence is already actively searched for by the Bir\'{e}fringence Magn\'{e}tique du Vide (BMV) \cite{Cadene:2013bva}, the Polarizzazione del Vuoto con Laser (PVLAS) \cite{DellaValle:2015xxa} and the Observing Vacuum with Laser (OVAL) \cite{Fan:2017fnd} experiments.
In these searches the probe photons are delivered by a continuous wave laser, and the birefringence signal is induced in a meter-sized magnetic field of a
few Tesla. The effective optical path-length for the probe photons in the magnetic field, and hence the number of polarization flipped signal photons, is increased substantially by the use of optical high finesse cavities.
To allow for a test of the QED prediction, the sensitivity of these ongoing searches still needs to be improved; for recent developments see \cite{Battesti:2018bgc} and references therein.

An alternative route to verify vacuum birefringence has been put forward by \cite{Heinzl:2006xc}, proposing to use a bright linearly polarized x-ray beam as probe and an optical high-intensity laser as pump; cf. also \cite{DiPiazza:2006pr,Dinu:2013gaa}.  
As the birefringence signal is inversely proportional to the wavelength of the probe and directly proportional to the number of photons available for probing, employing an XFEL as probe seems most promising, particularly given the great progress in x-ray polarimetry achieved in recent years \cite{Marx:2011,Marx:2013xwa}.
Such an experiment is scheduled, e.g., at the Helmholtz International Beamline for Extreme Fields at the European XFEL \cite{Schlenvoigt:2016}.
For proposals of vacuum birefringence experiments with dipole, synchrotron and gamma radiation, cf. \cite{Kotkin:1996nf,Nakamiya:2015pde,Ilderton:2016khs,King:2016jnl,Bragin:2017yau}.
Besides, also note the recently reported indications for the relevance of QED vacuum birefringence for the optical polarimetry of a neutron star \cite{Mignani:2016fwz,Capparelli:2017mlv,Turolla:2017tqt}, and \cite{Caiazzo:2018evl} for another discussion of vacuum birefringence effects in a cosmological context.

For other signatures of strong-field QED and further theoretical proposals to verify signatures of QED vacuum nonlinearities, see the reviews \cite{Dittrich:2000zu,Marklund:2008gj,Dunne:2008kc,Heinzl:2008an,DiPiazza:2011tq,Battesti:2012hf,King:2015tba,Karbstein:2016hlj,Inada:2017lop}, and references therein.

The standard approach to allow for an analytical estimate of the effect of QED vacuum birefringence is based on the determination of the refractive indices of the two physical probe photon propagation eigenmodes \cite{Toll:1952,Baier,BialynickaBirula:1970vy,Adler:1971wn} in infinitely extended constant electromagnetic fields. This determination explicitly makes use of the fact that due to translational invariance in such fields the four-momentum $k^\mu$ of traversing probe photons remains unaltered, allowing for the definition of global refractive indices depending on the strength and orientation of the background field.
In constant fields the difference of these refractive indices $\Delta n$ can be straightforwardly translated into a relative phase shift as $\Delta\phi=2\pi\frac{d}{\lambda_p}\Delta n$, where $d$ denotes the distance traveled by the probe photons in the constant electromagnetic field and $\lambda_p$ is the probe wavelength.
Now, conventional approximations aiming at analytical insights into the effect of vacuum birefringence in manifestly inhomogeneous fields employ a locally constant field approximation on the level of this formula, resulting in the simple {\it ad hoc} substitution of $\frac{d}{\lambda_p}\Delta n\to \frac{1}{\lambda_p}\int{\rm d}s\,\Delta n(s)$, where $s$ parameterizes the optical path of the probe photons in the pump field \cite{Heinzl:2006xc}.
The inhomogeneity of the pump results in a position $s$ dependent refractive index difference $\Delta n(s)$.
We emphasize that this derivation heavily relies on the assumption of translational invariance, which is of course manifestly violated in inhomogeneous fields.
In turn, effects generic to inhomogeneous fields, such as momentum exchanges between pump and probe fields are completely neglected from the outset \cite{Karbstein:2015qwa}, and potentially large deviations from more refined calculations are to be expected.

Having pointed out the relevance of the scattering of polarization-flipped signal photons outside the forward cone of the x-ray probe in all-optical studies of vacuum birefringence \cite{Karbstein:2015xra}, in \cite{Karbstein:2016lby} we have studied the phenomenon of vacuum birefringence in the collision of an XFEL probe and a near-infrared high-intensity pump laser pulse in unprecedented detail, based upon a reformulation of the effect of vacuum birefringence as a {\it vacuum emission process} \cite{Karbstein:2014fva}.
In this study, we modeled both lasers as linearly-polarized pulsed paraxial Gaussian beams and accounted for a finite collision angle as well as spatio-temporal misalignments of the beam foci.
The birefringence signal is maximum for counter propagating, perfectly aligned beams whose polarization vectors differ by an angle of $\frac{\pi}{4}$.

As it is usually the case when refining the theoretical description of an effect by accounting for more and more experimentally relevant parameters,
the more realistic modeling of the above vacuum birefringence scenario came along with more complicated expressions for the differential number of signal photons scattered into a perpendicular mode ${\rm d}^3N_\perp$ as compared to simplistic analytical estimates \cite{Heinzl:2006xc}. They generically involve integrations which cannot be performed analytically, and hence require the use of numerics for their evaluation.
The lack of an explicit representation of these results makes it hard to infer highly relevant characteristics, such as scaling laws describing how the signature changes under the variation of a given parameter.

In the present work, we aim at bridging this gap by providing new handy analytical expressions for the differential number of polarization flipped signal photons.
Starting point of our considerations are the results obtained in \cite{Karbstein:2016lby}.
More specifically, we do not focus on the most generic collision geometry as in \cite{Karbstein:2016lby}, but stick from the outset to the case of a strict counter-propagation geometry and the optimal choice for the polarization alignment of the beams.
At the same time, we still account for the effects of spatio-temporal misalignments of the beam foci; cf. Fig.~\ref{fig:scenario}.
Note, that the invoked simplifications also match the experimental situation: while the beam axes and polarizations can be aligned with sufficiently high precision, spatio-temporal offsets, e.g. due to beam jitter, generically occur and are typically hard to control in high-intensity laser experiments.

In a sense, the present paper amounts to a rather straightforward and logical continuation of \cite{Karbstein:2016lby}.
At the same time, it represents a significant step further by providing handy analytical expressions for the differential number of polarization flipped signal photons allowing for quantitative estimates in the planning of potential discovery experiments. 

Our paper is organized as follows:
In Sec.~\ref{sec:stateoftheart} we detail the starting point of our considerations and the involved approximations.
Section~\ref{sec:results} is devoted to the derivation of the new analytical expressions for the differential number of polarization flipped signal photons.
Moreover, some immediate applications of the obtained formulas are outlined.
Finally, we end with conclusions in Sec.~\ref{sec:conclusions}.

\section{Considered scenario}\label{sec:stateoftheart}

We assume both laser beams to be well-described as paraxial Gaussian beams, supplemented by Gaussian pulse profiles; for the field profiles, cf. Eqs.~(9) and (12) of \cite{Karbstein:2016lby}.
The QED vacuum induced effective interaction of the laser beams is treated on the level of a locally constant field approximation (LCFA) of the one-loop Heisenberg-Euler effective Lagrangian \cite{Heisenberg:1935qt,Schwinger:1951nm}.
As detailed, e.g., in \cite{Karbstein:2015cpa,Gies:2016yaa}, this approximation should be well-justified for weakly varying electromagnetic fields, characterized by photon energies much smaller than the electron rest energy $mc^2$.

\begin{figure}
\center
\includegraphics[width=0.9\textwidth]{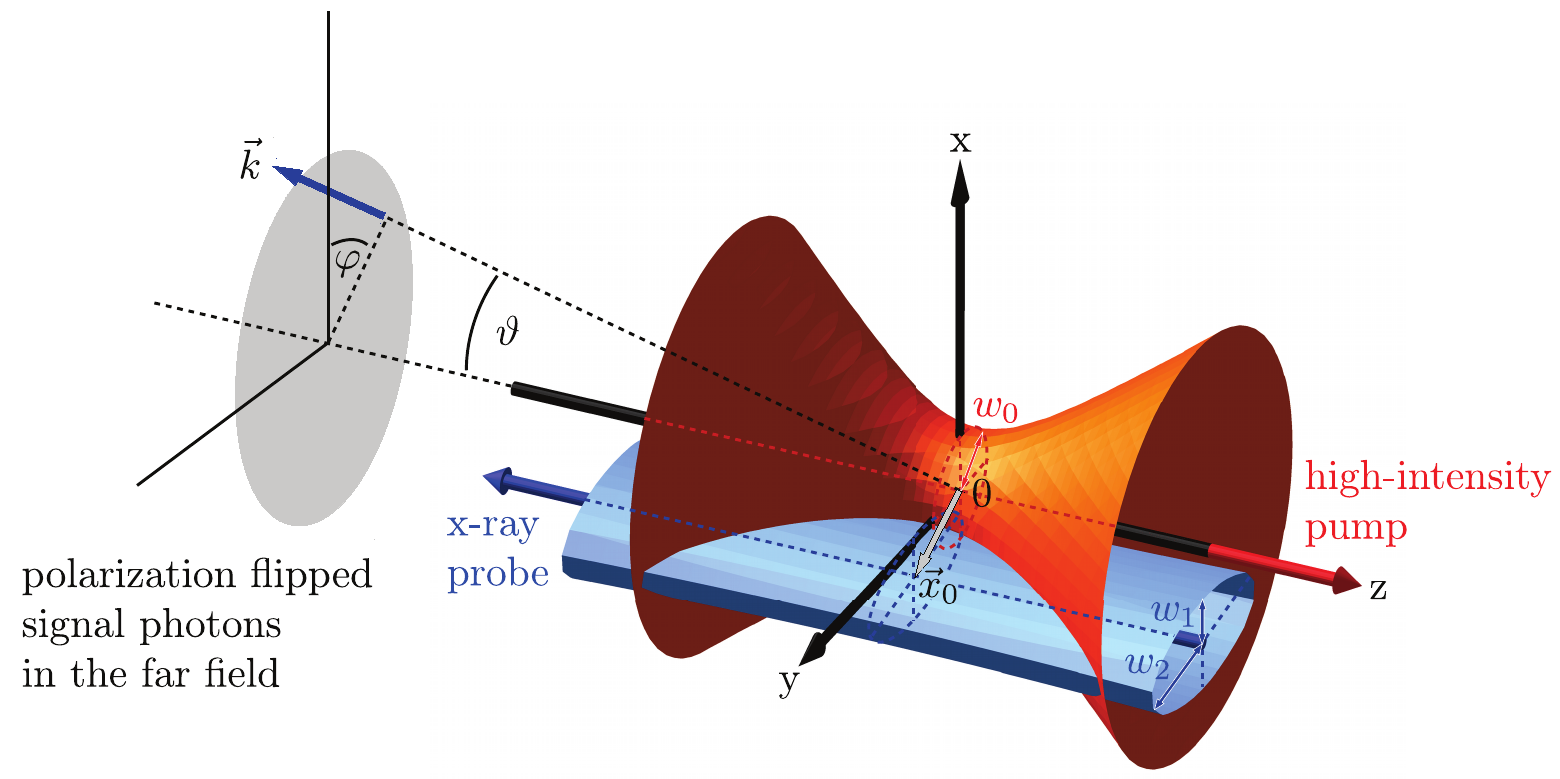}
\caption{Illustration of the scenario considered in this article. The high-intensity laser beam (beam waist $w_0$, beam focus at $\vec{x}=0$) exhibits a rotational symmetry about its beam axis and propagates along the positive $\rm z$ axis.
The XFEL probe propagates in negative $\rm z$ direction and has a generic elliptical transverse profile characterized by the two waist parameters $w_1$ and $w_2$. Its beam focus can be spatially offset from that of the high-intensity beam by $\vec{x}_0$.
The signal of vacuum birefringence is encoded in polarization flipped x-ray signal photons (wave vector $\vec{k}$), potentially deflected from the beam axis of the probe by an angle of $\vartheta\ll1$, and to be detected in the far field.}
\label{fig:scenario}
\end{figure}

The high-intensity laser beam (wavelength $\lambda$, waist $w_0$, Rayleigh range ${\rm z}_R=\pi w_0^2/\lambda$, pulse energy $W$ and pulse duration $\tau$) is assumed to exhibit a rotational symmetry around its beam axis. Without loss of generality we assume it to propagate in positive $\rm z$ direction. It is brought into collision with an XFEL pulse propagating in exactly opposite direction, i.e., in negative $\rm z$ direction.
As it maximizes the number of attainable signal photons, the counter-propagation geometry is also of most interest for experimental attempts aiming at the measurement of vacuum birefringence. 
Due to its comparably mild focusing, it is well-justified to adopt the {\it infinite Rayleigh range approximation} for the XFEL beam (wavelength $\lambda_p$, photon energy $\omega=2\pi/\lambda_p$, $N$ photons per pulse of duration $T$) in the interaction volume \cite{Karbstein:2015xra,Karbstein:2016lby}. More specifically, we allow for generic elliptical transverse XFEL beam profiles, characterized by two perpendicular waists $w_i$ with $i\in\{1,2\}$. Note, that for all considered cases, the associated Rayleigh ranges $\mathfrak{z}_{R,i}=\pi w_i^2/\lambda_p$ fulfill $\mathfrak{z}_{R,i}/{\rm z}_R=(w_i/w_0)^2 \lambda/\lambda_p \gg1$, justifying the use of the infinite Rayleigh range approximation.
For a graphical illustration of the considered scenario, cf. Fig.~\ref{fig:scenario}.

As mentioned in Sec.~\ref{sec:intro} above, for general collision geometries the expression for the differential number of polarization-flipped signal photons is rather unhandy and cannot be integrated analytically.
However, in the Appendix~A of Ref.~\cite{Karbstein:2016lby} we have shown that a substantial simplification is possible for the counter-propagation geometry, if one neglects the transverse widening $w({\rm z})=w_0\sqrt{1+({\rm z}/{\rm z}_R)^2}$ of the high-intensity laser beam radius with the longitudinal distance $\rm z$ from the beam focus at ${\rm z}=0$. To this end, one replaces $w({\rm z})\to w\geq w_0$ with a $\rm z$-independent effective beam radius $w$ in the interaction volume, while still accounting for the overall drop of its field strength $\sim w_0/w({\rm z})=1/\sqrt{1+({\rm z}/{\rm z}_R)^2}$ with increasing distance from the focus; cf. in particular Appendix~A of \cite{Karbstein:2016lby} where $w=w_{\rm eff}$ for the details.  
The latter substitution allows for a closed-form approximative expression for ${\rm d}^3N_\perp$ in terms of polynomials, exponentials and the error function, and is given in Eq.~(A2) of \cite{Karbstein:2016lby}. Choosing the {\it effective beam waist} $w$ adequately, this result amounts to an excellent approximation \cite{Karbstein:2016lby}:
it can be tuned such that the number of signal photons obtained with this approximation matches the corresponding outcome of a numerical evaluation of the full result presented in \cite{Karbstein:2016lby}, which is not employing the effective beam waist approximation.
Because of $w({\rm z})\geq w_0$, the naive identification $w=w_0$ reduces the strong-field volume and thus generically leads to an underestimation of the number of signal photons.
An alternative, especially convenient approximate choice for $w$, which does not need input from a numerical calculation, amounts to averaging $w({\rm z})$ over the beam's Rayleigh range, resulting in
\begin{equation}
 w=\frac{1}{{\rm z}_R}\int_{0}^{{\rm z}_R}{\rm dz}\,w({\rm z})=\frac{w_0}{2}[\sqrt{2}+{\rm arsinh}(1)]\simeq 1.15 w_0 \,.
 \label{eq:w}
\end{equation}
As demonstrated in Sec.~\ref{sec:results} below, the latter choice amounts to a reasonable approximation.
As it does not need any input from a numerical calculation, we will adopt it as our {\it approximation of choice}.

Unfortunately, the corresponding approximate expression for the differential number of polarization-flipped signal photons is still rather unhandy, and it is hard to infer the effect of changes of its various parameters on integrated numbers of signal photons.
It is the aim of the present work to overcome these limitations.
As they turn out to be substantially suppressed, we will in addition from the outset neglect contributions $\sim{\rm exp}\{-\frac{2\tau^2}{1+\frac{1}{2}(\frac{\tau}{T})^2}(\frac{\omega+{\rm k}}{8})^2\}$ in this expression -- or equivalently limit ourselves to the $q=+1$ contribution in Eq.~(A2) of \cite{Karbstein:2016lby} only.
In turn, in the Heaviside-Lorentz system and units where $c=\hbar=1$, the differential number of polarization flipped signal photons with momentum vector $\vec{k}={\rm k}(\cos\varphi\sin\vartheta,-\sin\varphi\sin\vartheta,-\cos\vartheta)$ in the far field (cf. Fig.~\ref{fig:scenario}) is given by
\begin{multline}
{\rm d}^3N_{\perp}
 \approx
m^4\frac{{\rm d}^3k}{(2\pi)^3}\,{\rm k}(w^2{\rm z}_R\tau)^2(1+\cos\vartheta)^2\,
\alpha\,\Bigl(\frac{\pi}{120}\Bigr)^2\Bigl(\frac{e{\mathfrak E}_0}{2m^2}\Bigr)^2\Bigl(\frac{e{\cal E}_0}{2m^2}\Bigr)^4\, \\
 \times\frac{(w_1w_2)^2}{w^4+2w^2(w_1^2+w_2^2)+4(w_1w_2)^2}\,\frac{1}{1+\frac{1}{2}(\frac{\tau}{T})^2}\, \\
 \times {\rm e}^{-\frac{1}{2}(w{\rm k}\sin\vartheta)^2\frac{w^2[w_1^2\cos^2(\varphi-\delta_0)+w_2^2\sin^2(\varphi-\delta_0)]+2(w_1w_2)^2}{w^4+2w^2(w_1^2+w_2^2)+4(w_1w_2)^2}}
 {\rm e}^{-4\frac{(w^2+2w_2^2)(\hat{\vec{a}}\cdot\vec{x}_0)^2
+(w^2+2w_1^2)(\hat{\vec{b}}\cdot\vec{x}_0)^2}{w^4+2w^2(w_1^2+w_2^2)+4(w_1w_2)^2}+\frac{8}{T^2}\frac{(2{\rm z}_R)^2-({\rm z}_0+t_0)^2}{1+\frac{1}{2}(\frac{\tau}{T})^2}} \\
\times{\rm e}^{-\frac{\tau^2}{1+\frac{1}{2}(\frac{\tau}{T})^2}(\frac{\omega-{\rm k}}{4})^2}\biggl|\sum_{\ell=\pm1} {\rm e}^{\ell{\rm z}_R\bigl[{\rm k}(1-\cos\vartheta)+\frac{8}{1+\frac{1}{2}(\frac{\tau}{T})^2}(\frac{\omega-{\rm k}}{4})-{\rm i}\frac{8}{T}\frac{2\frac{{\rm z}_0+t_0}{T}}{1+\frac{1}{2}(\frac{\tau}{T})^2}\bigr]} \\
 \times\biggl[1-{\rm erf}\biggl(\tfrac{\frac{4{\rm z}_R}{T}+\ell T(\frac{\omega-{\rm k}}{4})+\ell\frac{T{\rm k}}{8}(1-\cos\vartheta)[1+\frac{1}{2}(\frac{\tau}{T})^2]}{\sqrt{1+\frac{1}{2}(\frac{\tau}{T})^2}}
 -{\rm i}\ell\tfrac{2\frac{{\rm z}_0+t_0}{T}}{\sqrt{1+\frac{1}{2}(\frac{\tau}{T})^2}}\biggr)\biggr]\biggr|^2 \,, \label{eq:d3Nperpapprox}
\end{multline}
with spatial $\vec{x}_0=({\rm x}_0,{\rm y}_0,{\rm z}_0)$ and temporal $t_0$ offset parameters; $\alpha=\frac{e^2}{4\pi}\simeq\frac{1}{137}$ is the fine-structure constant.
Besides, $w_1$ is the waist of the XFEL probe in direction $\hat{\vec{a}}=(-\cos\delta_0,\sin\delta_0,0)$, and $w_2$ the waist in the perpendicular direction $\hat{\vec{b}}=(\sin\delta_0,\cos\delta_0,0)$.
The electric peak field amplitudes of the pump and probe lasers are ${\cal E}_0$ and ${\mathfrak E}_0$, respectively.
They can be related to the laser pulse parameters given above via \cite{Karbstein:2017jgh}
\begin{equation}
 {\cal E}_{0}^2\approx8\sqrt{\frac{2}{\pi}}\frac{W}{\pi w_0^2\tau} \quad\text{and}\quad \mathfrak{E}_{0}^2\approx8\sqrt{\frac{2}{\pi}}\frac{N\omega}{\pi w_1 w_2 T} \,. \label{eq:E2s}
\end{equation}
For completeness, note that these field amplitudes were underestimated in \cite{Karbstein:2016lby}, where inaccurate formulae relating the laser parameters and the peak field amplitudes were used:
To arrive at the correct values, all the results for the total signal photon numbers derived in \cite{Karbstein:2016lby} have to be rescaled by an overall factor of $(4\sqrt{2/\pi}/0.87)^3\simeq 49.37$.

\section{Results}\label{sec:results}

In Refs.~\cite{Karbstein:2015xra,Karbstein:2016lby} we have found that the spectrum of the signal photons is strongly peaked at the photon energy $\omega$ of the driving XFEL beam and quickly approaches zero towards both lower and higher frequencies within fractions of an electronvolt; cf., e.g., Fig.~3 of \cite{Karbstein:2015xra}.
Besides, the signal photons are predominantly emitted in the forward direction of the XFEL beam (corresponding to $\vartheta=0$), and their differential number decays rapidly with the azimuthal angle $\vartheta$ within a narrow angle regime fulfilling $\vartheta\ll1$ given that $\omega w_i\gg1$ for $i\in\{0,1,2\}$. The latter condition can be straightforwardly inferred from the argument of the exponential in the third line of \Eqref{eq:d3Nperpapprox},
and should be met for birefringence scenarios employing XFEL and high-intensity laser beams. For a more detailed discussion, cf. below.

In the following we will make use of these properties to arrive at a handier expression for the differential number of polarization-flipped signal photons.
More specifically, we devise the following procedure:
(i) Replace ${\rm k}$ by $\omega$ in slowly varying terms, accounting for the fact that $\vartheta\ll1$. In turn, all ${\rm k}$ in \Eqref{eq:d3Nperpapprox} apart from those in the three occurrences of the factor $(\frac{\omega-{\rm k}}{4})$ become $\omega$.
(ii) Expand all trigonometric functions of $\vartheta$ to leading order in $\vartheta\ll1$ and neglect higher-order terms.
(iii) Integrate over the signal photon energy ${\rm k}$, approximating the overall factor ${\rm k}^3$ contained in ${\rm dk}\,{\rm k}$ by $\omega^3$, and extending the integration boundaries to $\pm\infty$, respectively.
For convenience, we furthermore employ a substitution to the dimensionless integration variable $\kappa=\frac{\tau}{\sqrt{1+\frac{1}{2}(\frac{\tau}{T})^2}}(\frac{\omega-{\rm k}}{4})$ in the last step, and choose a coordinate system where $\delta_0=0$, such that $\hat{\vec{a}}\to-\hat{\vec{e}}_{\rm x}$ and $\hat{\vec{b}}\to\hat{\vec{e}}_{\rm y}$; cf. also Fig.~\ref{fig:scenario}. Of course, step (iii) is to be omitted if one is interested in the energy spectrum resolved signal photon numbers. However, from the overall factor $\sim\exp\{{-\frac{\tau^2}{1+\frac{1}{2}(\frac{\tau}{T})^2}(\frac{\omega-{\rm k}}{4})^2}\}$ in \Eqref{eq:d3Nperpapprox} we can immediately infer that the signal photon spectrum is strongly peaked at ${\rm k}=\omega$ and its $1/{\rm e}^2$ width is approximately given by $\Delta k\simeq\frac{8}{\tau}\sqrt{2+(\frac{\tau}{T})^2}$, which, e.g., for an experimentally realistic choice of $\tau=T=30\,{\rm fs}$ (cf. also below) yields $\Delta k\approx0.3\,{\rm eV}$.
In principle arising inelastic contributions due to QED vacuum fluctuation induced frequency mixing effects peaked at ${\rm k}\approx\omega\pm 2\Omega$ are substantially suppressed and thus have been neglected already from the outset; cf. \cite{Karbstein:2015xra,Karbstein:2016lby}. 

Implementing the steps (i)-(iii) in \Eqref{eq:d3Nperpapprox} and making use of the identities~\eqref{eq:E2s}, we obtain
\begin{align}
 \frac{{\rm d}^2N_{\perp}}{{\rm d}\varphi\,{\rm d}\vartheta\,\vartheta}
 &\approx
 \frac{N}{2\pi}\frac{4\alpha^4}{25(3\pi)^\frac{3}{2}}\,
\Bigl(\frac{W}{m}\frac{\lambdabar_{\rm C}}{w}\Bigr)^2\Bigl(\frac{w}{w_0}\frac{\omega}{m}\Bigr)^4\,\frac{\frac{w_1}{w}\frac{w_2}{w}}{1+2[(\frac{w_1}{w})^2+(\frac{w_2}{w})^2+2(\frac{w_1}{w}\frac{w_2}{w})^2]} \nonumber\\
 &\quad\times {\rm e}^{-\frac{1}{2}(\omega\vartheta)^2\frac{w_1^2\cos^2\varphi+w_2^2\sin^2\varphi+2w^2(\frac{w_1}{w}\frac{w_2}{w})^2}{1+2[(\frac{w_1}{w})^2+(\frac{w_2}{w})^2+2(\frac{w_1}{w}\frac{w_2}{w})^2]}}
 \nonumber\\
 &\quad\times {\rm e}^{-4\frac{[1+2(\frac{w_2}{w})^2](\frac{{\rm x}_0}{w})^2
+[1+2(\frac{w_1}{w})^2](\frac{{\rm y}_0}{w})^2}{1+2[(\frac{w_1}{w})^2+(\frac{w_2}{w})^2+2(\frac{w_1}{w}\frac{w_2}{w})^2]}}\,F\Bigl(\tfrac{\frac{4{\rm z}_R}{T}}{\sqrt{1+\frac{1}{2}(\frac{\tau}{T})^2}},\tfrac{2\frac{{\rm z}_0+t_0}{T}}{\sqrt{1+\frac{1}{2}(\frac{\tau}{T})^2}},\tfrac{T}{\tau}\Bigr) \,, \label{eq:d2N}
\end{align}
where $\lambdabar_{\rm C}=\hbar/(mc)$ is the reduced Compton wavelength of the electron, and we made use of the definition
\begin{align}
F(\chi,\chi_0,\rho)&:=\sqrt{\frac{1+2\rho^2}{3}}\,\chi^2\, {\rm e}^{2(\chi^2-\chi_0^2)} \nonumber\\
 &\quad\times \int_{-\infty}^\infty {\rm d}\kappa\,{\rm e}^{-\kappa^2}\,\biggl|\sum_{\ell=\pm1} {\rm e}^{2\ell(\rho\kappa-{\rm i}\chi_0)\chi} 
 \Bigl[1-{\rm erf}\Bigl(\ell(\rho\kappa
 -{\rm i}\chi_0)+\chi\Bigr)\Bigr]\biggr|^2 \,. \label{eq:F}
\end{align}

In strict forward direction, $\vartheta=0$, \Eqref{eq:d2N} scales as $\sim\omega^4$ with the probe photon energy.
For finite deflection angles $\vartheta$, also the $\omega^2$ dependence in the argument of the exponential in the second line of \Eqref{eq:d2N} becomes relevant.
A vanishing longitudinal offset ${\rm z}_0+t_0=0$ corresponds to $\chi_0=0$. In this case, all parameters in \Eqref{eq:F} are real valued.
Note, that all dependences of \Eqref{eq:d2N} on the pulse durations of the pump $\tau$ and probe $T$ pulses, as well as on the Rayleigh range ${\rm z}_R$ of the pump and the longitudinal offset ${\rm z}_0+t_0$ are encoded in the latter function.
For the parameters adopted in \cite{Karbstein:2016lby}: $\lambda=800\,{\rm nm}$, $\tau=T=30\,{\rm fs}$, ${\rm z}_0+t_0=0$ and $w_0=1\mu{\rm m}$, we have 
\begin{equation}
 F\Bigl(\sqrt{\tfrac{2}{3}}\tfrac{4{\rm z}_R}{T},0,1\Bigr)\simeq 1 \,, \label{eq:F1}
\end{equation}
implying that the presented formulas take a particularly compact form in this specific limit.

The decay of the differential number of signal photons with increasing impact parameter in transverse direction controlled by ${\rm x}_0$ and ${\rm y}_0$ is fully captured by the exponential function in the last line of \Eqref{eq:d2N}. It is independent of the integration variables and thus also describes the corresponding decay of the integrated numbers of signal photons.

It is often convenient to turn to {\it Cartesian angle coordinates}, defined as $X=\vartheta\cos\varphi$ and $Y=\vartheta\sin\varphi$, for which 
${\rm d}\varphi\,{\rm d}\vartheta\,\vartheta\to{\rm d}X{\rm d}Y$ and $w_1^2\cos^2\varphi+w_2^2\sin^2\varphi+2w^2(\frac{w_1}{w}\frac{w_2}{w})^2=w_1^2[1+2(\frac{w_2}{w})^2]X^2+w_2^2[1+2(\frac{w_1}{w})^2]Y^2$.
In these coordinates, the angular integrations to be performed to obtain $N_\perp$ from \Eqref{eq:d2N} reduce to elementary Gaussian integrals.
From \Eqref{eq:d2N} it is straightforward to infer the total number of polarization-flipped signal photons as
\begin{align}
\frac{N_{\perp}}{N}
 &\approx \frac{4\alpha^4}{25(3\pi)^\frac{3}{2}}\,
\Bigl(\frac{W}{m}\frac{\omega}{m}\Bigr)^2\Bigl(\frac{\lambdabar_{\rm C}}{w_0}\Bigr)^4\,\frac{1}{\sqrt{[1+2(\frac{w_1}{w})^2][1+2(\frac{w_2}{w})^2]}} \nonumber\\
 &\quad\times
 {\rm e}^{-4\frac{[1+2(\frac{w_2}{w})^2](\frac{{\rm x}_0}{w})^2
+[1+2(\frac{w_1}{w})^2](\frac{{\rm y}_0}{w})^2}{1+2[(\frac{w_1}{w})^2+(\frac{w_2}{w})^2+2(\frac{w_1}{w}\frac{w_2}{w})^2]}} \,
\underbrace{F\Bigl(\tfrac{\frac{4{\rm z}_R}{T}}{\sqrt{1+\frac{1}{2}(\frac{\tau}{T})^2}},\tfrac{2\frac{{\rm z}_0+t_0}{T}}{\sqrt{1+\frac{1}{2}(\frac{\tau}{T})^2}},\tfrac{T}{\tau}\Bigr)}_{\equiv F} \,, \label{eq:NperpbyN}
\end{align}
which scales as $\sim\omega^2$ with the probe photon energy.
As indicated in \Eqref{eq:NperpbyN}, to allow for a more compact representation of the respective equations, in the remainder we will often suppress the argument of the function $F$.

At the beginning of this section, we emphasized that the formulae presented here should be valid in the limit of $\omega w_i\gg1$, with $i\in\{0,1,2\}$.
Let us be a bit more specific what this restriction means for typical parameters available in experiment:
Assuming the XFEL photon energy to be given by $\omega=n\,{\rm keV}$ $\leftrightarrow$ $\lambda_p\approx\frac{1.2}{n}\,{\rm nm}$, with $n\geq1$, and the above criterion to be fulfilled sufficiently for $\omega w_i\geq10$, we obtain the condition $w_i\geq\frac{2}{n}\,{\rm nm}$ on the beam waists.
This condition is obviously met for optical high-intensity laser beams, for which $w_0={\cal O}(1)\mu{\rm m}$.
It imposes a more severe restriction on the XFEL waists $w_1$ and $w_2$.
However, all present and near-future XFEL facilities fulfill $w_i\gg\lambda_p$, i.e., cannot be focussed down to the diffraction limit, implying the above criterion to be satisfied as well.

The number of signal photons attainable for the particular case of the XFEL beam of a significantly smaller diameter than the optical high-intensity laser, i.e., $\{w_1,w_2\}\ll w_0$, nevertheless follow from the formal limit of $\{w_1,w_2\}\to0$ in \Eqref{eq:NperpbyN}.
In this case, \Eqref{eq:NperpbyN} is of a particularly simple form and reads
\begin{equation}
\frac{N_{\perp}}{N}\bigg|_{w_i=0}
 \approx \frac{4\alpha^4}{25(3\pi)^\frac{3}{2}}\,
\Bigl(\frac{W}{m}\frac{\omega}{m}\Bigr)^2\Bigl(\frac{\lambdabar_{\rm C}}{w_0}\Bigr)^4 {\rm e}^{-4\frac{{\rm x}_0^2+{\rm y}_0^2}{w^2}} F\Bigl(\tfrac{\frac{4{\rm z}_R}{T}}{\sqrt{1+\frac{1}{2}(\frac{\tau}{T})^2}},\tfrac{2\frac{{\rm z}_0+t_0}{T}}{\sqrt{1+\frac{1}{2}(\frac{\tau}{T})^2}},\tfrac{T}{\tau}\Bigr) ,
 \label{eq:NperpbyNpoint}
\end{equation}
and for vanishing spatio-temporal offset parameters $x_0^\mu=(t_0,\vec{x}_0)=0$,
\begin{equation}
\frac{N_{\perp}}{N}\bigg|_{w_i=x_0^\mu=0}
 \approx \frac{4\alpha^4}{25(3\pi)^\frac{3}{2}}\,
\Bigl(\frac{W}{m}\frac{\omega}{m}\Bigr)^2\Bigl(\frac{\lambdabar_{\rm C}}{w_0}\Bigr)^4 F\Bigl(\tfrac{\frac{4{\rm z}_R}{T}}{\sqrt{1+\frac{1}{2}(\frac{\tau}{T})^2}},0,\tfrac{T}{\tau}\Bigr) .
 \label{eq:NperpbyNpoint_nooff}
\end{equation}
For completeness, also note that further analytical insights are possible in the limit of very large probe pulse durations, significally larger than both the Rayleigh range ${\rm z}_R$ and the pulse duration $\tau$ of the high-intensity pump pulse.
To this end we consider the formal limit of $\{\frac{T}{\tau},\frac{T}{{\rm z}_R}\}\to\infty$.
Taking into account that $\lim_{T/\tau\to\infty}\{1-{\rm erf}(\frac{T}{\tau}\xi)\}=2\Theta(-\xi)$, where $\Theta(.)$ denotes the Heaviside function, it is straightforward to show that
\begin{equation}
 F\Bigl(\tfrac{\frac{4{\rm z}_R}{T}}{\sqrt{1+\frac{1}{2}(\frac{\tau}{T})^2}},0,\tfrac{T}{\tau}\Bigr)\Big|_{T\to\infty}\simeq \frac{\tau}{T}\sqrt{\frac{2\pi}{3}}\Bigl(\frac{8{\rm z}_R}{\tau}\Bigr)^2{\rm e}^{(\frac{8{\rm z}_R}{\tau})^2}\Bigl[1-{\rm erf}\bigl(\tfrac{8{\rm z}_R}{\tau}\bigr)\Bigr]. \label{eq:FTinfty}
\end{equation}
The overall scaling of \Eqref{eq:FTinfty} with $1/T$ agrees with our expectations: for plane wave probes of infinitely long pulse duration, the differential number of signal photons should be proportional to the probe photon current density $J\sim N/(\pi w_1w_2T)$, encoding the only dependence on $T$ \cite{Karbstein:2015cpa}.
Analogously, we can determine the limitting value of the same expression for a very large Rayleigh range of the pump, fulfilling ${\rm z}_{\rm R}\gg\{T,\tau\}$.
It is attainable from the formal limit of $\{\frac{{\rm z}_{\rm R}}{T},\frac{{\rm z}_{\rm R}}{\tau}\}\to\infty$, using the series expansion of the error function for large arguments.
The corresponding result is
\begin{equation}
 F\Bigl(\tfrac{\frac{4{\rm z}_R}{T}}{\sqrt{1+\frac{1}{2}(\frac{\tau}{T})^2}},0,\tfrac{T}{\tau}\Bigr)\Big|_{{\rm z}_{\rm R}\to\infty}\simeq \frac{4}{\sqrt{3\pi}}\,. \label{eq:FzRinfty}
\end{equation}
Moreover, noting that $F(\chi,0,1)=4\sqrt{\pi}\,\chi^2+{\cal O}(\chi^3)$, we can extract a limitting expression for $T=\tau$ and $\frac{{\rm z}_R}{\tau}\ll1$.
The latter is given by
\begin{equation}
  F\Bigl(\tfrac{\frac{4{\rm z}_R}{T}}{\sqrt{1+\frac{1}{2}(\frac{\tau}{T})^2}},0,\tfrac{T}{\tau}\Bigr)\Big|_{(\tau=T)\to\infty}\simeq \frac{128}{3}\sqrt{\pi}\,\Bigl(\frac{{\rm z}_R}{\tau}\Bigr)^2 \,, \label{eq:FTgleichtauinfty}
\end{equation}
where the $1/\tau^2$ dependence reflects the fact that the number of polarization flipped signal photons is proportional to ${\cal E}_0^4$; cf. \Eqref{eq:d3Nperpapprox}.

In Table~\ref{tab:vgl} we confront the total numbers of polarization-flipped signal photons $N_\perp$ normalized by the total number of incident XFEL photons for probing $N$ as obtained from \Eqref{eq:NperpbyN}, employing the identification~\eqref{eq:w}, with the corresponding results of a numerical integration of the full result for the differential number of polarization-flipped signal photons $N^\text{full}_\perp$ derived in \cite{Karbstein:2016lby} for various probe beam profiles.
This comparison is intended to obtain a feeling for the accuracy of the approximate result~\eqref{eq:NperpbyN}.
To this end, we adopt the laser pulse parameters of \cite{Karbstein:2016lby}: $W=30\,{\rm J}$, $\lambda=800\,{\rm nm}$, $\tau=T=30\,{\rm fs}$, $w_0=1\mu{\rm m}$, $\omega=12914\,{\rm eV}$, $N\simeq10^{12}$, as well as vanishing offset parameters.
Recall, that in this case $F\simeq1$; cf. \Eqref{eq:F1} above.
Table~\ref{tab:vgl} confirms that, even when sticking to the simple identification~\eqref{eq:w}, our analytical approximation~\eqref{eq:NperpbyN} allows for reasonable estimates of the attainable numbers of polarization-flipped signal photons.
For all cases shown in Table~\ref{tab:vgl} the relative deviations are below $\approx15\%$.
\begin{table}
\begin{tabular}{|c|c||c|c|c|}
 \hline
  $w_1/w_0$ & $w_2/w_0$ & $N_\perp/N$  & $N^\text{full}_\perp/N$ & $|1-N_\perp/N^\text{full}_\perp|$ \\
 \hline
 \hline
  $1/10$ & $1/10$ & $2.9\cdot10^{-11}$  & $3.0\cdot10^{-11}$ & $2.6\%$  \\
  $1/3$ & $1/3$ & $2.6\cdot10^{-11}$  & $2.6\cdot10^{-11}$ & $0.2\%$  \\
  $1$ & $1$ & $1.2\cdot10^{-11}$  & $1.1\cdot10^{-11}$ & $8.0\%$  \\
  $3$ & $3$ & $2.0\cdot10^{-12}$  & $1.8\cdot10^{-12}$ & $13.8\%$  \\
  $3$ & $1/10$ & $7.8\cdot10^{-12}$  & $7.4\cdot10^{-12}$ & $5.4\%$  \\
  $3$ & $1/3$ & $7.2\cdot10^{-12}$  & $6.8\cdot10^{-12}$ & $6.9\%$  \\
  $3$ & $1$ & $4.9\cdot10^{-12}$  & $4.4\cdot10^{-12}$ & $11.1\%$  \\
 \hline
 \end{tabular}
 \caption{Total numbers of polarization-flipped signal photons for different probe beam cross sections, characterized by the two waist parameters $w_1$ and $w_2$, measured in units of the pump waist $w_0$.
 The laser pulse parameters employed here are $W=30\,{\rm J}$, $\lambda=800\,{\rm nm}$, $\tau=T=30\,{\rm fs}$, $w_0=1\mu{\rm m}$, $\omega=12914\,{\rm eV}$, $N\simeq10^{12}$. In addition, we assume optimal collisions, i.e., $\vec{x}_0=0$ and $t_0=0$.
 Here, $N_\perp$ and $N^\text{full}_\perp$ denote our approximate results~\eqref{eq:NperpbyN} and the analogous outcome of a numerical integration of the full result for the differential number of polarization-flipped signal photons derived in \cite{Karbstein:2016lby}. The relative deviation of these quantities is measured by $|1-N_\perp/N^\text{full}_\perp|$.}
\label{tab:vgl}
\end{table}

Finally, we compare our result~\eqref{eq:NperpbyNpoint_nooff} for the total number of perpendicularly polarized signal photons attainable in the scenario where the XFEL beam fulfills $\{w_1,w_2\}\ll w_0$ and is perfectly aligned with the beam axis of the high-intensity pump with the simplistic estimation of the same quantity by \cite{Heinzl:2006xc}, based on the conventional approach to QED vacuum birefringence outlined in Sec.~\ref{sec:intro}.
By construction, this approach does not account for the angular spread of the signal photons, but only considers polarization-flipped signal photons propagating in exact forward direction.
In turn, only integrated signal photon numbers can be compared.
Identifying the peak intensity $I_0$ of the pump in the formulas given in \cite{Heinzl:2006xc} with the expression for ${\cal E}_0^2$ from \Eqref{eq:E2s}, their result can be expressed as
\begin{equation}
\frac{N_{\perp\text{\cite{Heinzl:2006xc}}}}{N}\bigg|_{w_i=x_0^\mu=0}=\Bigl(\frac{\Delta\phi}{2}\Bigr)^2
 \approx \frac{2048\alpha^4}{225\pi}\,
\Bigl(\frac{W}{m}\frac{\omega}{m}\Bigr)^2\Bigl(\frac{\lambdabar_{\rm C}}{w_0}\Bigr)^4\Bigl(\frac{{\rm z}_R}{\tau}\Bigr)^2 ,
 \label{eq:Heinzl}
\end{equation}
where $\Delta\phi$ denotes the relative phase shift between the two x-ray propagation eigenmodes in the high-intensity pump field; cf. \cite{Heinzl:2006xc} for the details.
In turn, the ratio of Eqs.~\eqref{eq:Heinzl} and \eqref{eq:NperpbyNpoint_nooff} is given by
\begin{equation}
 \Bigl(\frac{N_{\perp\text{\cite{Heinzl:2006xc}}}}{N}\Bigr)/\Bigl(\frac{N_{\perp}}{N}\Bigr)\bigg|_{w_i=x_0^\mu=0}\simeq 512\sqrt{\frac{\pi}{3}}\Bigl(\frac{{\rm z}_R}{\tau}\Bigr)^2/F\Bigl(\tfrac{\frac{4{\rm z}_R}{T}}{\sqrt{1+\frac{1}{2}(\frac{\tau}{T})^2}},0,\tfrac{T}{\tau}\Bigr). \label{eq:compare}
\end{equation}
For the specific laser pulse parameters given above, this results in a discrepancy as large as
\begin{equation}
 \Bigl(\frac{N_{\perp\text{\cite{Heinzl:2006xc}}}}{N}\Bigr)/\Bigl(\frac{N_{\perp}}{N}\Bigr)\bigg|_{w_i=x_0^\mu=0}\simeq 100. \label{eq:100}
\end{equation}
As argued above, given the rather crude {\it ad hoc} approximations involved in the derivation of \cite{Heinzl:2006xc}, a sizable deviation is not unexpected.
Admittedly, the observed discrepancy of roughly two orders of magnitude is rather impressive. It underpins even more the relevance of the new results presented here. 

To better understand the origin of this discrepancy it is worthwhile to have a closer look on the estimate of \cite{Heinzl:2006xc}:
Their derivation treats the probe as a monochromatic plane wave, formally implying the limit of $T\to\infty$.
Besides, the pulse duration of the pump is only indirectly accounted for via the pump intensity, requiring the condition $\frac{{\rm z}_R}{\tau}\ll1$ to hold.
These conditions are manifestly violated for the specific laser pulse parameters employed in \Eqref{eq:100}.
In turn, in the limit of $T=\tau$ and  $\frac{{\rm z}_R}{\tau}\ll1$ the ratio~\eqref{eq:compare} is expected to be closer to unity.
Upon insertion of \Eqref{eq:FTgleichtauinfty} into \Eqref{eq:compare}, we obtain
\begin{equation}
 \Bigl(\frac{N_{\perp\text{\cite{Heinzl:2006xc}}}}{N}\Bigr)/\Bigl(\frac{N_{\perp}}{N}\Bigr)\bigg|_{w_i=x_0^\mu=0}\simeq 4\sqrt{3}\approx 6.9\,. \label{eq:comparefair}
\end{equation}
Equation~\eqref{eq:comparefair} shows that even in the most favorable comparison, the total number of polarization flipped signal photons is substantially overestimated by \Eqref{eq:Heinzl}.

In the remainder of this article we briefly discuss some additional interesting results and scalings which can be straightforwardly inferred from our central result~\eqref{eq:d2N}.

Note, that in the far field the angular distribution of the photons constituting the XFEL probe beam should be well-described by \cite{Karbstein:2016lby}
\begin{equation}
 \frac{{\rm d}^2N}{{\rm d}\varphi\,{\rm d}\vartheta\, \vartheta}\simeq \frac{N}{2\pi}\,\omega^2 w_1 w_2
 \,{\rm e}^{-\frac{1}{2}(\omega\vartheta)^2(w_1^2\cos^2\varphi+w_2^2\sin^2\varphi)}\,, \label{eq:d2Nprobe}
\end{equation}
such that its radial $1/{\rm e}^2$ beam divergence is given by
\begin{equation}
 \theta(\varphi) = \frac{2}{\omega\sqrt{w_1^2\cos^2\varphi+w_2^2\sin^2\varphi}} \,. \label{eq:bdiv}
\end{equation}
Analogously, from \Eqref{eq:d2N} we can read off the radial beam divergence of the signal photons, resulting in
\begin{equation}
 \theta_\perp(\varphi) \simeq \theta(\varphi)\,\sqrt{1+2\Bigl[\Bigl(\frac{w_1}{w}\Bigr)^2+\Bigl(\frac{w_2}{w}\Bigr)^2+2\Bigl(\frac{w_1}{w}\frac{w_2}{w}\Bigr)^2\Bigr]} \frac{1}{\sqrt{1+\frac{2(\frac{w_1}{w}\frac{w_2}{w})^2}{(\frac{w_1}{w})^2\cos^2\varphi+(\frac{w_2}{w})^2\sin^2\varphi}}}\,.
\end{equation}
The latter result implies that the divergence of the signal photons is generically larger than that of the driving probe beam as $\{\frac{w_1}{w},\frac{w_2}{w}\}>0$.
In turn, apart from the asymptotic limit of an infinitesimal narrow probe beam profile described by $\{\frac{w_1}{w},\frac{w_2}{w}\}\to0$, for which the far-field divergences of the XFEL probe beam and of the signal photons agree with each other, the differential signal photons number~\eqref{eq:E2s} falls off slower with the azimuthal $\vartheta$ than the photons constituting the XFEL probe~\eqref{eq:d2Nprobe}.

Assuming the ratio of the differential numbers of polarization-flipped and inert probe photons in forward direction to be smaller than the polarization purity $\cal P$ of a given polarimeter, i.e., $\frac{{\rm d}N_\perp}{{\rm d}\varphi{\rm d}\vartheta \vartheta}/\frac{{\rm d}N}{{\rm d}\varphi{\rm d}\vartheta \vartheta}\bigl|_{\vartheta=0}<{\cal P}$, we can explicitly determine the polar angle $\vartheta=\vartheta_=(\varphi)$ for which
\begin{equation}
\Bigl(\frac{{\rm d}N_\perp}{{\rm d}\varphi{\rm d}\vartheta \vartheta}\Bigr)/\Bigl(\frac{{\rm d}N}{{\rm d}\varphi{\rm d}\vartheta \vartheta}\Bigr)\biggl|_{\vartheta=\vartheta_=(\varphi)}={\cal P}\,. \label{eq:theta=}
\end{equation}

At least in principle, signal photons scattered outside this angle should be directly measurable and no longer be background dominated; cf. also the detailed discussion in \cite{Karbstein:2016lby}.
Solving the latter equation for $\vartheta_=(\varphi)$, we find
\begin{align}
 \vartheta_{=}^2(\varphi)
&= \frac{1}{(\omega w)^2} \frac{1}{(\frac{w_1}{w}\frac{w_2}{w})^2-[(\frac{w_1}{w})^2\cos^2\varphi+(\frac{w_2}{w})^2\sin^2\varphi][(\frac{w_1}{w})^2+(\frac{w_2}{w})^2+2(\frac{w_1}{w}\frac{w_2}{w})^2]} \nonumber\\
&\quad\times\biggl\{\Bigl[1+2\Bigl(\frac{w_1}{w}\Bigr)^2+2\Bigl(\frac{w_2}{w}\Bigr)^2+4\Bigl(\frac{w_1}{w}\frac{w_2}{w}\Bigr)^2\Bigr] \nonumber\\
&\hspace*{1.5cm}\times\ln\biggl(\frac{4\alpha^4}{25(\sqrt{3\pi})^3}\Bigl(\frac{W}{m}\frac{\omega}{m}\Bigr)^2\Bigl(\frac{\lambdabar_{\rm C}}{w_0}\Bigr)^4
 \frac{1}{1+2[(\frac{w_1}{w})^2+(\frac{w_2}{w})^2+2(\frac{w_1}{w}\frac{w_2}{w})^2]}\frac{F}{{\cal P}}\biggr) \nonumber\\
 &\hspace*{1cm}-4\Bigl[1+2\Bigl(\frac{w_2}{w}\Bigr)^2\Bigr]\Bigl(\frac{{\rm x}_0}{w}\Bigr)^2-4\Bigl[1+2\Bigl(\frac{w_1}{w}\Bigr)^2\Bigr]\Bigl(\frac{{\rm y}_0}{w}\Bigr)^2\biggr\}\,.
\end{align}
With this result and \Eqref{eq:d2N} above, it is straightforward to determine the differential number $\frac{{\rm d}N_{\perp >}}{{\rm d}\varphi}\simeq\int_{\vartheta_=(\varphi)}^\infty{\rm d}\vartheta\,\vartheta\,\frac{{\rm d}^2N_{\perp}}{{\rm d}\varphi{\rm d}\vartheta \vartheta}$ of polarization-flipped signal photons emitted into directions fulfilling $\vartheta\geq\vartheta_=(\varphi)$, yielding
\begin{align}
 \frac{{\rm d}N_{\perp >}}{{\rm d}\varphi}
 &\approx
 \frac{N}{2\pi}{\cal P}\,
\frac{\frac{w_1}{w}\frac{w_2}{w}[1+2(\frac{w_1}{w})^2+2(\frac{w_2}{w})^2+4(\frac{w_1}{w}\frac{w_2}{w})^2]}{(\frac{w_1}{w})^2\cos^2\varphi+(\frac{w_2}{w})^2\sin^2\varphi+2(\frac{w_1}{w}\frac{w_2}{w})^2} \nonumber\\
 &\quad\times \biggl(\frac{2\alpha^2}{5(3\pi)^{\frac{3}{4}}}\frac{W}{m}\frac{\omega}{m} \Bigl(\frac{\lambdabar_{\rm C}}{w_0}\Bigr)^2
 \frac{\sqrt{F/{\cal P}}}{\sqrt{1+2[(\frac{w_1}{w})^2+(\frac{w_2}{w})^2+2(\frac{w_1}{w}\frac{w_2}{w})^2]}}\biggr)^{\kappa}
 \nonumber\\
 &\quad\times {\rm e}^{-2\frac{[1+2(\frac{w_2}{w})^2](\frac{{\rm x}_0}{w})^2
+[1+2(\frac{w_1}{w})^2](\frac{{\rm y}_0}{w})^2}{(\frac{w_1}{w})^2+(\frac{w_2}{w})^2+2(\frac{w_1}{w}\frac{w_2}{w})^2-(\frac{w_1}{w}\frac{w_2}{w})^2/[(\frac{w_1}{w})^2\cos^2\varphi+(\frac{w_2}{w})^2\sin^2\varphi]}} \,, \label{eq:dN>}
\end{align}
with
\begin{equation}
 \kappa= \frac{1+2[(\frac{w_1}{w})^2+(\frac{w_2}{w})^2+2(\frac{w_1}{w}\frac{w_2}{w})^2]}{(\frac{w_1}{w})^2+(\frac{w_2}{w})^2+2(\frac{w_1}{w}\frac{w_2}{w})^2-(\frac{w_1}{w}\frac{w_2}{w})^2/[(\frac{w_1}{w})^2\cos^2\varphi+(\frac{w_2}{w})^2\sin^2\varphi]}\,.
\end{equation}
It is instructive to compare \Eqref{eq:dN>} with the analogous result for the differential number of signal photons integrated over all $\vartheta$,
\begin{align}
 \frac{{\rm d}N_{\perp}}{{\rm d}\varphi}
 &\approx
 \frac{N}{2\pi}\frac{4\alpha^4}{25(3\pi)^\frac{3}{2}}\,
\Bigl(\frac{W}{m}\frac{\omega}{m}\Bigr)^2\Bigl(\frac{\lambdabar_{\rm C}}{w_0}\Bigr)^4\,F \nonumber\\
 &\quad\times \frac{\frac{w_1}{w}\frac{w_2}{w}}{(\frac{w_1}{w})^2\cos^2\varphi+(\frac{w_2}{w})^2\sin^2\varphi+2(\frac{w_1}{w}\frac{w_2}{w})^2}
 \nonumber\\
 &\quad\times {\rm e}^{-4\frac{[1+2(\frac{w_2}{w})^2](\frac{{\rm x}_0}{w})^2
+[1+2(\frac{w_1}{w})^2](\frac{{\rm y}_0}{w})^2}{1+2[(\frac{w_1}{w})^2+(\frac{w_2}{w})^2+2(\frac{w_1}{w}\frac{w_2}{w})^2]}} \,, \label{eq:dN}
\end{align}
with regard to the scaling of the differential numbers of signal photons with the probe photon energy $\omega$.
While the latter is proportional to an overall factor of $\omega^2$ independent of $\varphi$, the former exhibits a $\varphi$ dependent scaling, reaching its extremal values for $\varphi=0$ (direction of half-axis $w_1$) and $\varphi=\frac{\pi}{2}$ (direction of $w_2$).
More specifically, we infer a scaling $\sim\omega^{\beta_i}$ with exponential $\beta_1=2+(\frac{w}{w_1})^2$ for $\varphi=0$ and $\beta_2=\beta_1|_{w_1\leftrightarrow w_2}$ for $\varphi=\frac{\pi}{2}$.
Note, that $\beta_i>2$ for all possible choices of $w_1$ and $w_2$.

For circular transverse probe profiles $w_1=w_2$ it is even possible to perform the integration over $\varphi$ in \Eqref{eq:dN>} analytically, yielding
\begin{align}
 \frac{N_{\perp >}}{N}\bigg|_{w_1=w_2}
 &\approx
 {\cal P}\,
\Bigl[1+2\Bigl(\frac{w_1}{w}\Bigr)^2\Bigr] \biggl(\frac{2\alpha^2}{5(3\pi)^{\frac{3}{4}}}\frac{W}{m}\frac{\omega}{m}\Bigl(\frac{\lambdabar_{\rm C}}{w_0}\Bigr)^2
 \frac{\sqrt{F/{\cal P}}}{1+2(\frac{w_1}{w})^2}\biggr)^{2+(\frac{w}{w_1})^2}
 \nonumber\\
 &\quad\times {\rm e}^{-2(\frac{w}{w_1})^2[(\frac{{\rm x}_0}{w})^2
+(\frac{{\rm y}_0}{w})^2]} \,. \label{eq:Ncirc}
\end{align}

Finally, we note that our analytical formulas can also be employed to obtain an estimate of the impact of a background of larger radial divergence in the far field on the signal-to-background ratio of the perpendicularly polarized signal photons.
To this end, we assume the far-field angular distribution of the $N$ photons constituting the XFEL probe beam to be made up of two contributions, containing (i) $\bar{N}$ and (ii) $\delta N$ x-ray photons, respectively. In turn, we have $N=\bar N+\delta N$.
More specifically, we describe both of these contributions by \Eqref{eq:d2Nprobe}, employing (i) the single substitution $N\to\bar{N}$, and (ii) both substitutions $N\to\delta N$ and $\omega\to\epsilon\omega$.
Here, the parameter $\epsilon\ll1$ is introduced to make the distribution wider, while at the same time reducing its peak amplitude. 
In order to model a background of radial divergence $\theta(\varphi)/\epsilon$, cf. \Eqref{eq:bdiv} above,
at the level of $b\ll1$ for the differential XFEL photon number in the far field, we set $\delta N=N/(1+\epsilon^2/b)\ll N$, which immediately implies $\epsilon^2/b\gg1$ and $\bar N=N(\epsilon^2/b)/(1+\epsilon^2/b)$.
In the formal limit of $\epsilon^2/b\to\infty$, we have $\bar N\to N$ and $\delta N\to 0$.

As an immediate consequence of their distinct radial divergences, for small (larger) values of $\vartheta$ the differential photon number is dominated by the contribution proportional to $\bar N$ ($\delta N$).
In turn, for low enough background levels the condition~\eqref{eq:theta=} is now even met twice, namely for $\vartheta^{(i)}_=(\varphi)\approx\vartheta_=(\varphi)|_{F\to(1+b/\epsilon^2)F}$ and
\begin{align}
 \vartheta^{(ii)}_=(\varphi)&\approx
 \sqrt{\frac{(\frac{w_1}{w}\frac{w_2}{w})^2-[(\frac{w_1}{w})^2\cos^2\varphi+(\frac{w_2}{w})^2\sin^2\varphi][(\frac{w_1}{w})^2+(\frac{w_2}{w})^2+2(\frac{w_1}{w}\frac{w_2}{w})^2]}{(\frac{w_1}{w}\frac{w_2}{w})^2-[(\frac{w_1}{w})^2\cos^2\varphi+(\frac{w_2}{w})^2\sin^2\varphi]\{\epsilon^2[(\frac{w_1}{w})^2+(\frac{w_2}{w})^2+2(\frac{w_1}{w}\frac{w_2}{w})^2]+\frac{\epsilon^2-1}{2}\}}}\nonumber\\
 &\quad\times\vartheta_=(\varphi)\big|_{F\to(\frac{1}{\epsilon^2}+\frac{1}{b})F}\,.
\end{align}
As in the idealized situation scenario detailed above, due to the faster decay of the differential number of XFEL probe photons relative to the differential signal photon number, from a certain polar angle $\vartheta(\varphi)=\vartheta^{(i)}_=(\varphi)$ onwards, the perpendicularly polarized photon signal should no longer be background dominated. 
However, due to presence of an additional background exhibiting a much wider radial divergence than the polarization flipped signal photons, this regime is now delimited from above by the angle $\vartheta(\varphi)=\vartheta^{(ii)}_=(\varphi)$. For larger polar angles the signal photons are again background dominated. 
In turn, the differential number of perpendicularly polarized signal photons fulfilling the criterion $\frac{{\rm d}N_\perp}{{\rm d}\varphi{\rm d}\vartheta \vartheta}/\frac{{\rm d}N}{{\rm d}\varphi{\rm d}\vartheta \vartheta}\bigl|_{\vartheta=0}\geq{\cal P}$ is given by $\frac{{\rm d}N_{\perp >}}{{\rm d}\varphi}\simeq\int_{\vartheta_=^{(i)}(\varphi)}^{\vartheta_=^{(ii)}(\varphi)}{\rm d}\vartheta\,\vartheta\,\frac{{\rm d}N_{\perp}}{{\rm d}\varphi}$.

\section{Conclusions}\label{sec:conclusions}

In this  for novel insights into the scaling of the effect with the waists and pulse durations of both laser beams, as well as the Rayleigh range of the high-intensity beam.
Moreover, they account for the effects of transverse offsets of the beam axes of the XFEL and high-intensity laser beams, and non-optimal focus overlaps in longitudinal direction.
As an immediate application, our results allow for the determination of an analytical expression for the angular divergence of the perpendicularly polarized signal photons.
This observable could previously be extracted only from fits to numerical data points; cf. \cite{Karbstein:2015cpa,Karbstein:2016lby}.

We expect our formulas to be very useful for the planning and optimization of experimental scenarios aiming at the detection of vacuum birefringence in XFEL/high-intensity laser setups, such as the one put forward at HIBEF at the European XFEL. In a second step, the optimal scenarios for experiment identified on the basis of the approximate results for the polarization-flipped signal photon numbers provided in the present article can then be analyzed quantitatively in a full numerical analysis along the lines of \cite{Karbstein:2016lby}.

\acknowledgments

I would like to thank Holger~Gies, Elena~A. Mosman, Gerhard~G.~Paulus, Kai~Sven~Schulze, Chantal~Sundqvist, Ingo~Uschmann and Matt~Zepf for collaboration and stimulating discussions on the presented topic.

\end{document}